\documentclass{article}
\usepackage{amssymb,amsmath,cite}
\newcommand{\angstrom}{\textup{\AA}}
\usepackage{graphicx}
\usepackage{hyperref}
\usepackage{epstopdf}
\date{}
\title {\textbf{Time- and memory-efficient representation of complex mesoscale potentials}}
\author {Grigory Drozdov, Igor Ostanin and Ivan Oseledets}
\begin{document}
\maketitle

\begin{abstract}

We apply the modern technique of approximation of multivariate functions - tensor train cross approximation - to the problem of the description of physical interactions between complex-shaped bodies in a context of computational nanomechanics. In this note we showcase one particular example - van der Waals interactions between two cylindrical bodies - relevant to modeling of carbon nanotube systems. The potential is viewed as a tensor (multidimensional table) which is represented in compact form with the help of tensor train decomposition. The described approach offers a universal solution for the description of van der Waals interactions between complex-shaped  nanostructures and can be used within the framework of such systems of mesoscale modeling as recently emerged mesoscopic distinct element method (MDEM).

\end{abstract}

\section*{Introduction}

Mesoscale coarse-grained mechanical modeling \cite{Buehler2006, Volkov2010, Ostanin2013, Ostanin2014} is an important tool for \textit{in silico} characterization of nanostructures. The typical coarse-grained modeling approach is based on the idea of the representation of a nanostructure (\textit{e.g.} nanotube, nanoparticle, \textit{etc.}) as a set of elements, which are treated either as point masses or solid bodies (rigid or deformable), interacting via bonding or non-bonding potentials. Compared to regular molecular dynamic (MD) approaches, such models are easily as scalable as MD, while rendering much higher computational efficiency. For example, they can be used for accurate mechanical modeling of the representative volume elements of complex nanostructured materials. However, such models require the efficient and reliable description of van der Waals (vdW) interaction potentials between complex-shaped elements. Such potentials depend on multiple parameters. For example, the interaction between two rigid bodies of a fixed size requires tabulation of six-dimensional array with good precision. The description of interactions between the bodies of variable size, which is required for adaptive models, or interactions between deformable elements, requires even more degrees of freedom and leads to larger multidimensional arrays of data. This makes simple tabulation of such potentials impractical, since the time to construct such a table and the memory to store it grows exponentially with the number of dimensions (the problem known as ``curse of dimensionality''). The approach that is often used in the field is to design relatively simple analytical approximations for such interaction potentials, representing the potentials sought as separable functions \cite{Volkov2010, Ostanin2013}. Such an approach may be efficient in some situations, but in many cases it leads to over-simplifications and incorrect mechanics on the larger scale.
In this work we suggest the reliable and universal approach for storing arbitrarily complex mesoscale potentials as compressed multidimensional tables (following the terminology accepted in data science and computational linear algebra communities, such tables will be referred to as tensors below). The compression is reached via tensor-train cross approximation \cite{Oseledets2011, Oseledets2010}. It allows to construct the approximation of a tensor with the desired precision, using relatively small number of explicitly computed potential values. Such approaches were applied earlier in the field of computational chemistry \cite{Baranov2015}, however, they are still widely unknown to the community of multiscale mechanical modeling.
In order to demonstrate the pipeline of tensor approximation in application to mesoscale models, we consider relatively simple yet pratically important problem of the vdW interaction between two equal-sized cylinders. This problem is important in a context of mesoscale mechanical modeling of large assemblies of carbon nanotubes (CNTs). As has been showed earlier, the use of simple pair potentials for the description of intertube interactions between cylindrical CNT segments leads to significant artifacts in model's behavior. In the previous works \cite{Volkov2010, Ostanin2013} the problem was addressed with sophisticated analytical approximations. However, these approximations can not be transparently generalized onto more complex situations  - CNTs of different diameters, curved CNT segments \textit{etc.} Our approach presented here does not suffer from such a lack of generality and can be used in a number of similar problems of the description of interactions between complex-shaped nanostructures.

\section*{Method}

In this section we describe our technique of construction of the compressed table of the interaction potential between complex-shaped bodies with the example of vdW interactions between two equal-sized cylindrical segments of CNTs. Following the coarse-graining approach developed in \cite{Ostanin2013, Ostanin2014}, we idealize CNT segments as interacting cylindrical surfaces of uniform density. Total vdW interaction between two segments is found via the integration of standard Lennard-Jones (LJ) potential:

\begin{equation}
u_{LJ}(r)=4\varepsilon\left( \left(
\frac{\sigma}{r}\right)^{12}-\left(
\frac{\sigma}{r}\right)^{6}\right)
\end{equation}

where $r$ is the distance between particles, $\sigma\sqrt[6]{2}$ is the equilibrium distance, $-\varepsilon$ is the energy at the equilibrium distance. For carbon-carbon interaction, we accept $\sigma=3.851\ \angstrom$, $\varepsilon=0.004\ eV$. In order to avoid the integration of an artificial singular part of LJ potential, we replace the potential in near-singular region $(0,r_0), r_0 = 3.8 \angstrom $, with the cubic spline $u(r)$, satisfying $u(0)=3\ eV, u'(0) = 0, u(r_{0}) = u_{LJ}(r_{0}), u'(r_{0}) = u'_{LJ}(r_{0}) $.

We assume that carbon atoms are uniformly distributed over
surfaces of cylinders with the surface density
$\rho=\frac{4}{3\sqrt3a_{C-C}}$, where
$a_{C-C}=1.42\angstrom$ is the equilibrium carbon-carbon
bond length. The potential between cylindrical segments of
nanotubes is then represented as the integral over the
surface of each cylinder:

\begin{equation}
\label{integral}
U_t=\int\limits_{S_1}{\int\limits_{S_2}{\rho^2u_{LJ}\left(r\right)dS_1dS_2}}
\end{equation}
where $S_1$ and $S_2$ are cylinders side surfaces, $dS_1$
and $dS_2$ are the elements of surfaces, $r$ is the
distance between $dS_1$ and $dS_2$. The shapes of function
$r$ for different parametrizations are given in the Appendix. 
In order to describe the mutual position and orientation of two cylinders one needs four independent variables - six variables for general rigid bodies are reduced by two due to axial symmetries of the cylinders. Since the choise and order of these four variables is important for the approximation technique we intend to use, we compare two different parameterizations. The first one includes one distance and three angles ($R,  \alpha_1,  \alpha_2,  \alpha_{12} $ ( Fig. 1(a))), whereas the second utilizes three distances and one angle ( $ t_1, t_2, H, \gamma  $(Fig. 1(b))). For both choices of independent variables, we specify the regular grid in four-dimensional space with the appropriate tabulation limits. We sample $n$ points along each independent variable, which results in $n^4$ tabulated potential values.

\begin{figure}[h]
	\centering
	\includegraphics[height=0.2\textheight]{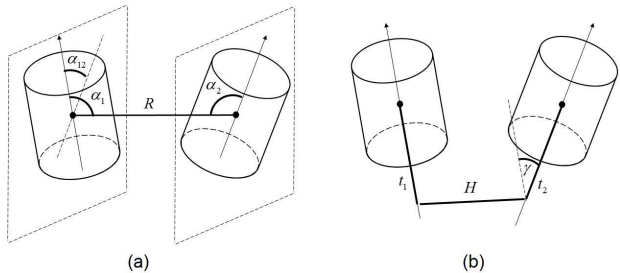}
	\protect\caption{\label{1 param} Two possible parameterizations of mutual position of two cylinders}
\end{figure}

Clearly, it is impossible to perform such expensive calculations ``on the fly'' for millions of interacting bodies. Therefore, within the direct approach we have to tabulate this potential function on multidimensional dense grid. It is possible for our rather simple example, but becomes computationally prohibitive in terms of required time and memory for more complex structures and corresponding potentials. Below we describe a general approach that is capable to compute such multidimensional tables fast and store them in a compact form.

Our approach is based on tensor train decomposition (TT) and tensor train cross approximation \cite{Oseledets2011, Oseledets2010}. In the TT format each element of $d$-dimensional tensor $U\left(i_1,i_2,\ldots,i_d\right)$ is
represented as the product of $d$ matrices:
\begin{equation}
\label{TT}
U\left(i_1,i_2,\ldots,i_d\right)=\sum\limits_{\alpha_0,\alpha_1,\ldots,\alpha_d}G_1\left(\alpha_0,i_1,\alpha_1\right)G_2\left(\alpha_1,i_2,\alpha_2\right)\ldots
G_d\left(\alpha_{d-1},i_d,\alpha_d\right)
\end{equation}
Here indices $\alpha_k$ are changing from $1$ to $r_k$,
which are called TT-ranks. Each matrix
$G_k\left(\alpha_k-1,i_k,\alpha_k\right)$ has the size
$r_{k-1}\times r_k$, depends on only one index of the original tensor, and these
matrices form the three-dimensional tensors, which are
called cores of the TT-decomposition. $r_0$ and $r_d$
are assigned to $1$. Intermediate TT ranks $r_k$ are determined by the TT approximation procedure to provide given approximation precision $\epsilon_{tt}$ and are conditioned by the structure of the approximated tensor.

Since tensors generated by the discretization of physically meaningful functions are typically ether low-rank or well approximated by the low-rank tensors, the decomposition (\ref{TT}) provides a powerful tool to represent such tensors in a compact form. The technique of construction of the best approximation in TT format with a given precision is described in \cite{Oseledets2010}. In this work we use the standard package TT toolbox for construction of the low-rank approximation of our potential.
In application to mesoscale modeling, TT decomposition can be used in two different ways. The first way is to accelerate the design of a full potential table and the second - to store it in a compact form. In the latter case the calculation of a single value of the potential in a nodal point would require computation of $d$ matrix-vector products, with matrix dimensions  $r_{k-1}\times r_k$.

\section*{Results and discussion}

We constructed TT decompositions for a tensor of
cylinder-cylinder interactions, for few different values of
mesh refinement and accuracy. The potential of interaction between two cylindrical segments of (10,10) CNTs with the radius $R_c = 6.78 \angstrom$  and height $2 R_c = 13.56 \angstrom$  was considered. The four-dimensional
integral (\ref{integral}) was computed using regular
rectangle quadratures over azimuthal angles of cylinders,
and Gauss quadratures over cylinders axial directions. The
number of integration points was chosen adaptively to
provide the desired integration precision $\epsilon_i = 10^{-3}$. In order to keep
our computations fast, we restricted ourselves with the best
integration precision $\epsilon_i = 10^{-3}$. In our case of
four-dimensional tensor TT-decomposition contains two
matrices and two three-dimensional tensors as TT-cores. The tensor indices correspond to the sampling of the continuous potential function on a regular grid: $U(i_1,i_2,\ldots,i_d)=U_t(x_1, x_2,\ldots, x_d)$, where 
$x_d = x_d^{min} + i/n \cdot  (x_d^{max} - x_d^{min} )$. The tabulation limits $(x_d^{min}$, $x_d^{max})$ for both parametrizations are given in Table \ref{table1}.

\begin{table}[h!]
	\caption{\label{table1} Tabulation limits for two paremetrizations used.}  \centering
	\begin{tabular}{|l|l|l|l|}
		\hline 
		\multicolumn{2}{c|}{Parametrization 1} &
		\multicolumn{2}{c}{Parametrization 2}\\
		
		\hline  
		
		\multicolumn{1}{c|}{$R$} & \multicolumn{1}{c|}{$(2  R_c, 5  R_c)$} & \multicolumn{1}{c|}{$H$} & \multicolumn{1}{c}{$(2  R_c, 5  R_c)$}  \\
		
		\hline  
		
		\multicolumn{1}{c|}{$\alpha_1$} & \multicolumn{1}{c|}{$(0, \pi)$} & \multicolumn{1}{c|}{$t_1$} & \multicolumn{1}{c}{$(-5  R_c, 5  R_c)$} \\
		
		\hline  
		
		\multicolumn{1}{c|}{$\alpha_2$} & \multicolumn{1}{c|}{$(0, \pi)$} & \multicolumn{1}{c|}{$t_2$} & \multicolumn{1}{c}{$(-5  R_c, 5  R_c)$}  \\
		
		\hline  
		
		\multicolumn{1}{c|}{$\alpha_{12}$} & \multicolumn{1}{c|}{$(0, \pi)$} & \multicolumn{1}{c|}{$\gamma$} & \multicolumn{1}{c}{$(0, \pi)$} \\
		
		\hline
	\end{tabular}
\end{table}

We
can define the compression of the tensor as the ratio of the
numbers of elements in the initial tensor and in all the
cores of the resulting TT-decomposition. The TT compression
was performed for two different parameterizations of
cylinders mutual position and different orders of variables
(TT decomposition is not invariant with respect to
permutation of tensor indices, our results are given for TT decompositions with the permutation of variables providing the lowest ranks of TT cores and the best compression). Tables \ref{table2} and
\ref{table3} give the compression number obtained for the
first and second parametrizations respectively, as a
function of the grid refinement (number of points per each
dimension) and the requested approximation precision $\epsilon_{tt}$. It
appears that even black-box application of the TT algorithm
gives significant compression of the original tensor (which
also roughly corresponds to the time needed to construct
this tensor approximation).

\begin{table}[h!]
\caption{\label{table2} The compression for the first
parameterization} \centering
\begin{tabular}{|l|l|l|l|l|}
\hline \multicolumn{1}{c|}{} & \multicolumn{4}{c}{TT approximation precision $\epsilon_{tt}$} \\
\hline \multicolumn{1}{c|}{} &
\multicolumn{2}{c|}{$10^{-3}$} &
\multicolumn{2}{c}{$10^{-2}$}\\
\hline \multicolumn{1}{c|}{$n$} & \multicolumn{1}{c|}{Max
rank} & \multicolumn{1}{c|}{Compression, $\%$} &
\multicolumn{1}{c|}{Max rank} &
\multicolumn{1}{c}{Compression, $\%$}\\
\hline \multicolumn{1}{c}{$10$} & \multicolumn{1}{c}{$20$}
& \multicolumn{1}{c}{$29.4$} & \multicolumn{1}{c}{$13$} &
\multicolumn{1}{c}{$18.2$}\\
\hline \multicolumn{1}{c}{$20$} & \multicolumn{1}{c}{$41$}
& \multicolumn{1}{c}{$13.6$} & \multicolumn{1}{c}{$16$} &
\multicolumn{1}{c}{$3.4$}\\
\hline \multicolumn{1}{c}{$30$} & \multicolumn{1}{c}{$47$}
& \multicolumn{1}{c}{$5.3$} & \multicolumn{1}{c}{$16$} &
\multicolumn{1}{c}{$1$}\\
\hline \multicolumn{1}{c}{$40$} & \multicolumn{1}{c}{$48$}
& \multicolumn{1}{c}{$2.4$} & \multicolumn{1}{c}{$16$} &
\multicolumn{1}{c}{$0.4$}\\
\hline
\end{tabular}
\end{table}

\begin{table}[h!]
\caption{\label{table3} The compression for the second
parameterization} \centering
\begin{tabular}{|l|l|l|l|l|}
\hline \multicolumn{1}{c|}{} & \multicolumn{4}{c}{TT approximation precision $\epsilon_{tt}$} \\
\hline \multicolumn{1}{c|}{} &
\multicolumn{2}{c|}{$10^{-3}$} &
\multicolumn{2}{c}{$10^{-2}$}\\
\hline \multicolumn{1}{c|}{$n$} & \multicolumn{1}{c|}{Max
rank} & \multicolumn{1}{c|}{Compression, $\%$} &
\multicolumn{1}{c|}{Max rank} &
\multicolumn{1}{c}{Compression, $\%$}\\
\hline \multicolumn{1}{c}{$10$} & \multicolumn{1}{c}{$27$}
& \multicolumn{1}{c}{$44.8$} & \multicolumn{1}{c}{$19$} &
\multicolumn{1}{c}{$30$}\\
\hline \multicolumn{1}{c}{$20$} & \multicolumn{1}{c}{$65$}
& \multicolumn{1}{c}{$24.7$} & \multicolumn{1}{c}{$33$} &
\multicolumn{1}{c}{$10.2$}\\
\hline \multicolumn{1}{c}{$30$} & \multicolumn{1}{c}{$85$}
& \multicolumn{1}{c}{$12.1$} & \multicolumn{1}{c}{$38$} &
\multicolumn{1}{c}{$3.4$}\\
\hline \multicolumn{1}{c}{$40$} & \multicolumn{1}{c}{$96$}
& \multicolumn{1}{c}{$6.3$} & \multicolumn{1}{c}{$39$} &
\multicolumn{1}{c}{$1.5$}\\
\hline
\end{tabular}
\end{table}

\begin{figure}[h!]
\centering
\includegraphics[height=0.2\textheight]{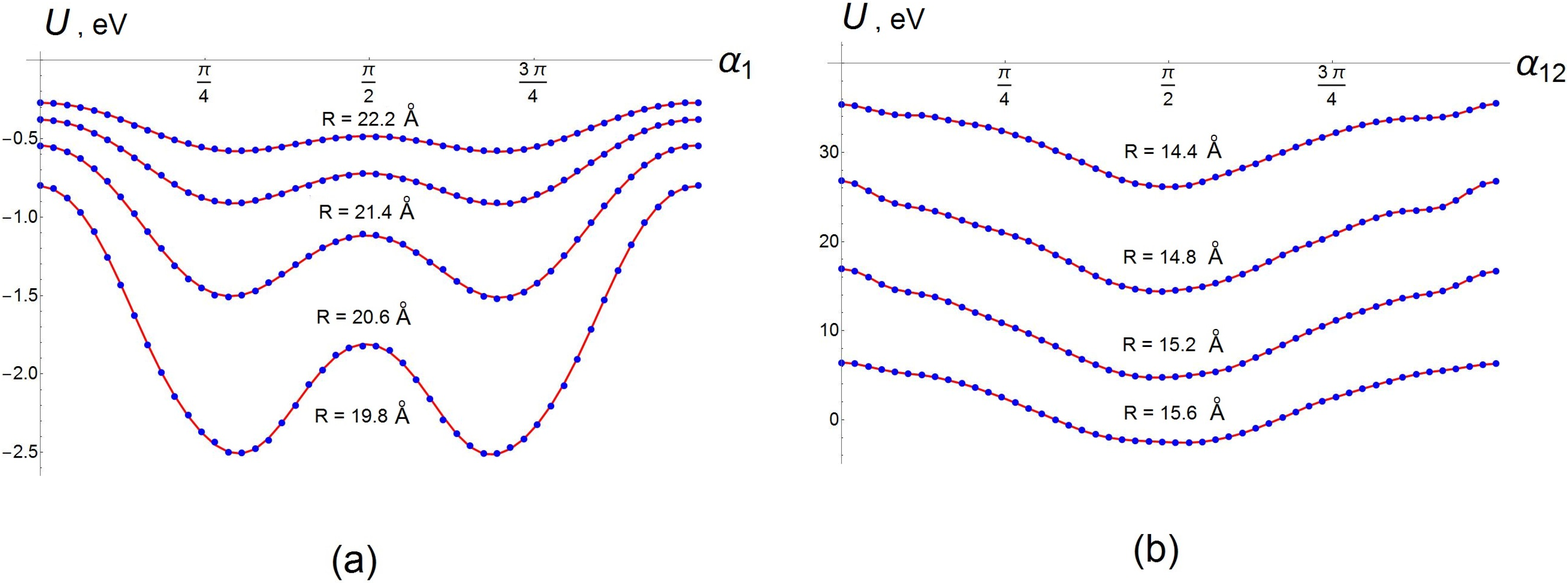}
\protect\caption{\label{slices} The potential in the first parametrization ($U( R, \alpha_1, \alpha_2, \alpha_{12} )$, solid red) and its TT approximation (dotted blue, $n=50$, $\epsilon_{tt}= 10^{-2}$), as a function of one of parametrization angles and few different values of $R$: (a) $\alpha_1 \in (0, \pi)$, $\alpha_2 = 0$, $\alpha_{12}=0$, $R = 19.8, 20.6, 21.4, 22.2$, (b) $\alpha_{12} \in (0, \pi)$, $\alpha_1 = 0$, $\alpha_2=0$, $R = 14.4, 14.8, 15.2, 15.6$.}  
\end{figure}

Figure \ref{slices} illustrates the quality of the approximation of a complex potential relief at few cross-sections that correspond to in-plane (a) and out-of-plane (b) rotaton of one cylinder with respect to another for different intercenter distances. As we can see, the compressed potential representation fully reproduce original tensor.

\section*{Conclusion}

In this note we have demonstrated the successful application of the technique that have recently emerged in the community of computational multilinear algebra - tensor-train cross approximation - to the problem of representation of the interaction potentials in modern mesoscale models. Within this approach the multidimensional potential table can be constructed based on just few direct calculations of the potential values, which dramatically accelerates construction of such a table. In the case when memory efficient representation is required, this table can be stored in a compact form and then each required value of the potential can be recovered at a moderate computational cost. In case when performance of on-the-fly computations is the highest priority, our approach can be used for fast construction of the full multidimensional table. The instruments for such approximaiton are freely available as components of  the open source software TT-toolbox (\url{https://github.com/oseledets/ttpy}),  and our code, performing the approximation of cylinder-cylinder vdW interaction potential, can be found at \url{https://bitbucket.org/iostanin/cnt_potential_tt_compression/}. The resulting approximations can be easily stored in either full or compact form, and further used in the framework of such existing mesoscale models as MDEM \cite{Ostanin2013, Ostanin2014, Ostanin2015}. Our approach appears to be particularly useful for the problems involving interactions between complex deformable shapes - \textit{e.g.} CNT segments with variable length, radius, curvature and flatness.

\section*{Acknowledgements}

Authors gratefully acknowledge partial financial support from Russian Foundation for Basic Research under grants RFBR 16-31-00429, RFBR 16-31-60100. This work was partially supported by an Early Stage Innovations grant from NASA’s Space Technology Research Grants Program.

\section*{Appendix}

Here we give the explicit shapes for the functions $r$ for two parameterizations used. In both cases the distance depends on four variables describing the mutual position of two cylinders, and four more variables - axial coordinates $z_1, z_2$ and azimuthal angles $ \varphi_1, \varphi_2 $, describing the positions of surface elements $dS_1$ and $dS_2$ on cylinders' side surfaces.
First parameterization:
$$r(dS_1,dS_2)=r( R, \alpha_1, \alpha_2, \alpha_{12}, z_1,z_2,\varphi_1,\varphi_2)=$$
$$\Big(\big(z_2
\sin{\alpha_{12}}\sin{\alpha_2}-R_c\left(\cos{\alpha_2}
\cos{\varphi_2}\sin{\varphi_2}+\sin{\varphi_1}\right)\big)^2+$$
$$+\big(R\sin{\alpha_1}+z_2\left(\cos{\alpha_2}\sin{\alpha_1}
-\cos{\alpha_1}\cos{\alpha_{12}}\sin{\alpha_2}\right)+$$
$$+R_c\big(\cos{\varphi_2}\sin{\alpha_1}\sin{\alpha_2}+
\cos{\alpha_1}\cos{\alpha_{12}}\cos{\alpha_2}\cos{\varphi_2}-$$

$$\cos{\alpha_1}\sin{\alpha_{12}}\sin{\varphi_2}-\cos{\varphi_1}
\big)\big)^2+$$
$$+\big(R\cos{\alpha_1}+z_2\left(\cos{\alpha_1}\cos{\alpha_2}+
\cos{\alpha_{12}}+\cos{\alpha_{12}}\sin{\alpha_1}
\sin{\alpha_2}\right)+$$
$$+R_c\big(\cos{\alpha_1}\cos{\varphi_2}\sin{\alpha_2}-
\cos{\alpha_{12}}\cos{\alpha_2}\cos{\varphi_2}
\sin{\alpha_1}+ $$
$$\sin{\alpha_1}\sin{\alpha_{12}}\sin{\varphi_2}
\big)-z_1\big)^2\Big)^{1/2}$$

Second parameterization:
$$r(dS_1,dS_2)=r(t_1, t_2, H, \gamma, z_1,z_2,\varphi_1,\varphi_2)=$$
$$\Big(
\big(\left(t_2+z_2\right)\sin{\gamma}+R_c\left(\sin{\varphi_1}-
\cos{\gamma}\sin{\varphi_2}\right)\big)^2+$$
$$+\big(H+R_{c}\left(\cos{\varphi_2}-\cos{\varphi_1}
\right)\big)^2+$$
$$+\big(\left(t_2+z_2\right)\cos{\gamma}-R_c\sin{\gamma}\sin{\varphi_2}-
\left(t_1+z_1\right)\big)^2\Big)^{1/2}$$

\end{document}